\documentclass[sigplan, 9pt]{acmart}

\usepackage{xspace}
\usepackage{multirow}
\usepackage{enumitem}

\newcommand{\namex}{ConfigSpec\xspace}

\title{\namex: Profiling-Based Configuration Selection for Distributed Edge--Cloud Speculative LLM Serving}

\author{Xiangchen Li}
\affiliation{%
  \institution{Virginia Tech}
  \city{Blacksburg, Virginia}
  \country{USA}
}
\email{lixiangchen@vt.edu}

\author{Saeid Ghafouri}
\affiliation{%
  \institution{Queen's University Belfast}
  \city{Belfast}
  \country{Northern Ireland, UK}
}
\email{s.ghafouri@qub.ac.uk}

\author{Jiakun Fan}
\affiliation{%
  \institution{Virginia Tech}
  \city{Blacksburg, Virginia}
  \country{USA}
}
\email{jiakunfan@vt.edu}

\author{Babar Ali}
\affiliation{%
  \institution{Queen's University Belfast}
  \city{Belfast}
  \country{Northern Ireland, UK}
}
\email{b.ali@qub.ac.uk}

\author{Hans Vandierendonck}
\affiliation{
  \institution{Queen's University Belfast}
  \city{Belfast}
  \country{Northern Ireland, UK}
}
\email{h.vandierendonck@qub.ac.uk}

\author{Dimitrios S. Nikolopoulos}
\affiliation{%
  \institution{Virginia Tech}
  \city{Blacksburg, Virginia}
  \country{USA}
}
\email{dsn@vt.edu}

\acmYear{2026}\copyrightyear{2026}
\setcopyright{cc}
\setcctype[4.0]{by}
\acmConference[TDIS '26]{4th International Workshop on Testing Distributed Internet of Things Systems}{April 27--30, 2026}{Edinburgh, Scotland Uk}
\acmBooktitle{4th International Workshop on Testing Distributed Internet of Things Systems (TDIS '26), April 27--30, 2026, Edinburgh, Scotland Uk}
\acmDOI{10.1145/3802513.3803483}
\acmISBN{979-8-4007-2608-8/26/04}


\begin{document}

\begin{abstract}
Speculative decoding enables collaborative Large Language Model (LLM) inference across cloud and edge by separating lightweight token drafting from heavyweight verification. While prior systems show performance and cost benefits, practical deployment requires navigating a large configuration space spanning draft model variants, quantisation levels, speculative lengths, and heterogeneous edge devices. This paper presents \namex, a configuration-selection framework for distributed speculative LLM serving. \namex profiles edge devices and draft--target alignment, and models drafting throughput, acceptance rate, and power to evaluate goodput, verification cost efficiency, and energy efficiency across the joint configuration space. \textcolor{black}{Our analysis across three edge platforms and two LLM families reveals structurally conflicting optima. Firstly, goodput is maximised by the smallest, fastest draft model at device-dependent speculative lengths ($K^*{=}2$--10). Secondly, both cost and energy efficiency converge to $K{=}2$ due to a dominant bonus-token effect---with cost favouring the largest drafter for its high acceptance rate and energy favouring the smallest for its low power draw}. These conflicts confirm that no single fixed configuration can simultaneously optimise all objectives, underscoring the need for profiling-based configuration selection in disaggregated edge--cloud LLM inference.

\end{abstract}

\maketitle

\section{Introduction}
Deploying Large Language Models (LLMs) closer to end users has become increasingly attractive due to the need to reduce end-to-end latency, limit bandwidth usage \cite{tianclone}, and preserve data locality and privacy \cite{han2025privacy}. Many interactive, streaming, and Internet of Things (IoT) applications in the fields of healthcare, industrial IoT, surveillance, etc., benefit from near-source inference execution, where responsiveness and data sensitivity are critical, enabling LLM inference on edge and mobile platforms
\cite{yu2024edge}. For on-device inference, a wide range of model efficiency techniques have been explored, including quantization \cite{xiao2023smoothquant}, pruning \cite{frantar2023sparsegpt}, and architectural simplification, which aim to reduce computation, memory footprint, and energy consumption on resource-constrained hardware. While these techniques can substantially lower inference cost and enable partial deployment of LLM models on edge devices, they face fundamental limits, as aggressive compression often leads to accuracy degradation and diminishing returns \cite{chen2025efficientqat}. Consequently, fully executing high-capacity LLMs on edge devices alone remains impractical for many workloads, motivating hybrid cloud--edge execution models in which IoT devices offload sensory insights to edge, where edge devices perform lightweight or partial inference while delegating expensive computation to centralized infrastructure \cite{li2025adaptive}.

\textcolor{black}{Speculative decoding is a decoding paradigm for autoregressive language models in which a lightweight draft model proposes multiple tokens ahead of time, and a higher-capacity target model subsequently verifies them against the target distribution. The target model accepts correct prefixes and discards the remaining, which eliminates divergence likelihood} \cite{leviathan2023fast}. Crucially, speculative decoding preserves the output distribution of the target model despite using an auxiliary draft model \cite{leviathan2023fast,chen2023accelerating}. This process decouples generation into a drafting phase that prioritizes speed and a verification phase that enforces correctness \cite{leviathan2023fast,miao2023specinfer}. By enabling multiple tokens to be proposed and validated in a single step, speculative decoding exposes opportunities to overlap computation, amortize verification cost, and restructure inference across heterogeneous resources.

Speculative decoding can substantially improve throughput, system capacity, and cost efficiency in heterogeneous edge environments \cite{li2025sled, li2026wisp, xu2024edgellm}. By distributing generation across devices with different computational capabilities, these systems achieve better utilization of both edge and server resources while maintaining output quality. However, speculative decoding exposes a fundamental challenge in practice that is orthogonal to mechanism design, namely configuration selection. Designers must decide which draft model variants to deploy on what edge devices, \textcolor{black}{and how to set the speculative decoding length,} choices that directly impact performance, resource efficiency, monetary cost, and energy consumption across heterogeneous hardware. In realistic deployments, these configurations interact in non-trivial ways with hardware constraints, workload characteristics, and service-level objectives. Larger draft models typically improve token acceptance rates, but consume more compute and memory on the device, potentially limiting concurrency and degrading energy efficiency. No single configuration dominates across devices, workloads, and optimization objectives.

Selecting configurations for distributed speculative LLM serving is challenging because system behavior depends on the interaction between draft model family, model size, quantization level, edge hardware, \textcolor{black}{speculative decoding length,} and target model alignment. These factors jointly determine verified-token throughput, verification cost efficiency, and energy efficiency, and their effects are often non-intuitive. For example, increasing draft model size improves acceptance rate but may reduce goodput and increase energy per verified token; similarly, platform differences can dominate model-size differences. As a result, configuration decisions cannot be \textcolor{black}{inferred} from model accuracy or hardware capability alone, and require systematic profiling and cross-metric evaluation.

In this paper, we present \namex, a framework that combines systematic device profiling, draft--target alignment measurement, and analytical performance modeling to evaluate configuration quality. \namex characterizes each draft model and edge platform using measurable quantities---drafting throughput, acceptance rate against a target model, and device power draw---and \textcolor{black}{analytically factors cloud verification effects into a set of parameters,} mapping these profiles to deployment-relevant metrics including goodput, verification cost efficiency, and energy efficiency. By grounding configuration evaluation in measurable primitives while preserving the structural properties of speculative decoding, \namex enables fast, repeatable comparison of model families, quantization levels, and edge platforms prior to large-scale deployment.

In summary, this paper makes the following contributions:

\begin{itemize}
    \item We identify configurations---specifically the joint choice of draft model, \textcolor{black}{speculative decoding length,} and edge platform---as a primary determinant of performance, cost, and energy in distributed edge--cloud speculative LLM serving.
    \item We design \namex, a profiling-based configuration evaluation framework that measures drafting throughput, draft--target acceptance rate, and device power, and \textcolor{black}{systematically maps them to system goodput, cost, and device energy}.
    \item Exhaustive enumeration of the joint $(M,Q,K)$ configuration space reveals structurally conflicting optima: goodput favours smallest draft at device-dependent $K^*{=}2$--$10$, while both cost and energy converge to $K{=}2$ via a bonus-token effect---yet cost selects the \emph{largest} drafter and energy the \emph{smallest}. These conflicts produce trade-offs of up to $2.9\times$ in goodput, $2.2\times$ in cost, and $7.8\times$ in energy between objective-optimal configurations on same device, confirming that no single fixed setting can simultaneously optimise all metrics.
\end{itemize}

\section{Related Work}
Related works are categorized into three groups: speculative decoding algorithms, speculative decoding systems for edge and distributed serving, and edge–cloud collaboration for LLM inference.

\textbf{Speculative Decoding Algorithms.}
Speculative decoding was originally introduced as a lossless acceleration technique for autoregressive generation, enabling multiple tokens to be proposed in parallel by a lightweight draft model and verified by a larger target model. Leviathan et al.~\cite{leviathan2023fast} formalize speculative decoding and prove that it preserves the output distribution of standard decoding while achieving substantial speedups without retraining or architectural changes. Subsequent work extends speculative decoding to new regimes and workloads. MagicDec~\cite{sadhukhanmagicdec} demonstrates that speculative decoding remains effective for moderate to long sequences and high-throughput inference, identifying bottleneck shifts with batch size and proposing adaptive drafting strategies to improve throughput and latency.

\textbf{Speculative Decoding Systems for Edge and Distributed Serving.}
Recent systems integrate speculative decoding into practical LLM serving architectures. EdgeLLM~\cite{xu2024edgellm} applies speculative decoding to on-device inference, addressing memory constraints by combining draft models with efficient branch navigation, adaptive fallback, and compute–IO pipelining. SLED~\cite{li2025sled} reinterprets speculative decoding as a mechanism for collaborative edge computing, allowing heterogeneous edge devices to draft tokens locally while a shared server verifies them in batches. WISP~\cite{li2026wisp} further identifies wasted drafting time and verification interference as key scalability bottlenecks in distributed speculative serving and proposes dynamic drafting and SLO-aware batching to improve system capacity and goodput.


\textbf{Edge--Cloud Collaboration for LLM Inference.}
Beyond speculative decoding, several works study collaborative inference across edge and cloud resources. CLONE~\cite{tianclone} explores algorithm–hardware co-design for latency-aware LLM inference on edge devices, relying on profiling and deployment-driven evaluation to optimize performance and energy efficiency. EdgeShard~\cite{zhang2024edgeshard} proposes partitioning LLMs into shards distributed across collaborative edge devices and cloud servers, optimizing device selection and partitioning decisions under heterogeneity and bandwidth constraints. While these approaches demonstrate the benefits of collaboration between edge and cloud, they do not model speculative decoding or address configuration selection across draft variants and speculative lengths.
\vspace{-0.15in}
\section{System Framework}

\begin{figure}[htbp]
    \centering
    \includegraphics[width=0.45\textwidth]{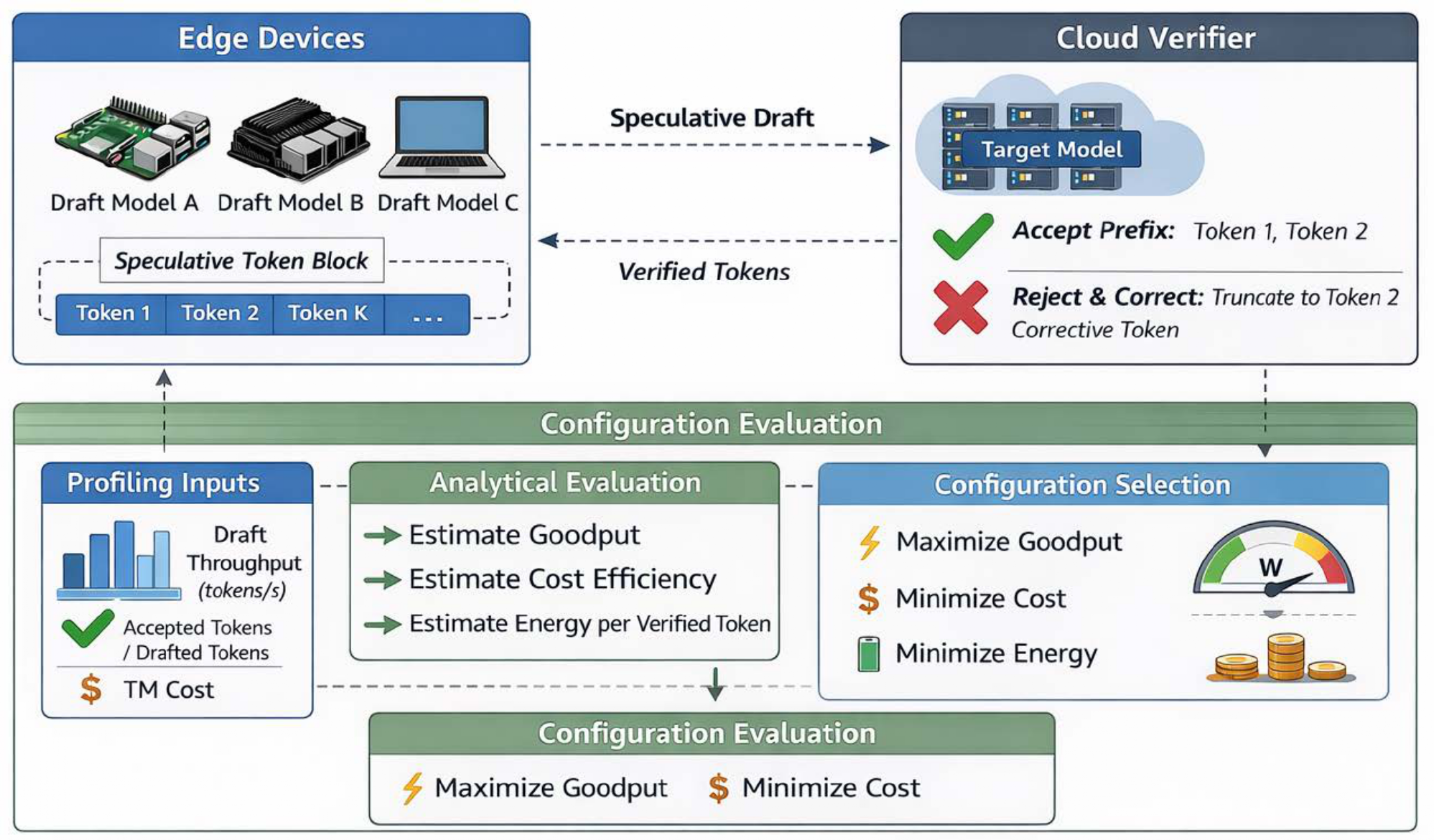}
    \caption{System abstraction used by \namex. Heterogeneous edge devices generate speculative tokens using local draft models and interact with a centralized cloud verifier.}
    \label{fig:system-overview}
    \vspace{-0.15in}
\end{figure}

\namex provides a configuration-profiling framework for distributed speculative LLM serving across heterogeneous edge platforms, as illustrated in Figure~\ref{fig:system-overview}. In the deployment model we consider, multiple edge devices execute inference requests using local draft models, while a centralized verifier hosts a target model that enforces correctness under speculative decoding semantics~\cite{leviathan2023fast}. \textcolor{black}{\namex evaluates configurations through profiling combined with an analytical model of speculative decoding rounds.
}

\subsection{Framework}

\paragraph{Serving abstraction.}
An inference request consists of autoregressive token generation. On each edge device, a draft model proposes tokens locally and submits them for verification by the target model. The verifier accepts the longest valid prefix and produces a corrective token upon divergence. Only accepted tokens contribute to final output.

\paragraph{Configuration profiling.}
For each draft model and edge platform, \namex profiles: (i) drafting throughput $v_d$, (ii) draft–target acceptance rate $\alpha(K)$, (iii) device power draw $P$.

\paragraph{Configuration space and selection.}
A configuration consists of a target model at the verifier, a draft model variant, and an edge platform. Given the measured profile, \namex evaluates configurations under objectives such as maximizing goodput, minimizing verification cost per token, or minimizing edge device energy per verified token.
\subsection{Performance, Cost, and Energy Model}

\textcolor{black}{
We model distributed speculative decoding using the same round abstraction employed in our evaluation. In each speculative round, an edge device drafts $K$ candidate tokens at rate $v_d$ (tokens/s), which are verified by a cloud-hosted target model with acceptance rate $\alpha(K)$ and verification latency $T_{\text{verify}}$.
}

\textcolor{black}{
\paragraph{Goodput.}
We define the \emph{goodput} as the verified token throughput under the speculative decoding round model:
}

\textcolor{black}{
\begin{equation}
  G(K) = \frac{K \cdot \alpha(K) + 1}{K / v_d + T_{\text{verify}}}
  \quad [\text{tok/s}],
  \label{eq:cost_efficiency}
\end{equation}
}

\textcolor{black}{
where the numerator counts the expected accepted tokens per round ($K\alpha(K)$ from the draft plus one bonus token at the first rejection), and the denominator is the round latency comprising local drafting time $K/v_d$ and remote verification latency $T_{\text{verify}}$.
}

\textcolor{black}{
\paragraph{Verification Cost Efficiency.}
Under a token-priced billing model with unit price $p$ (\$/token), each speculative round processes approximately $K$ tokens at the verifier. The cost efficiency in accepted tokens per dollar is:
}

\textcolor{black}{
\begin{equation}
  \eta_{\text{cost}} =
  \frac{K \cdot \alpha(K) + 1}{K \cdot p}
  = \frac{\alpha(K) + 1/K}{p}
  \quad [\text{tok/\$}],
\end{equation}
}

which depends only on the acceptance rate $\alpha(K)$, speculative length $K$, and verifier price $p$. \textcolor{black}{In practice, continuous batching and parallel prefill/decode can influence this cost. Accounting for the existing conflicting objectives and diverse search space, we considered token-priced billing for decoding without batching.}

\textcolor{black}{
\paragraph{Energy Efficiency.}
Let $P$ denote the average power draw of the edge device during drafting. Since verification occurs in the cloud, only local drafting time contributes to on-device energy consumption. The energy per verified token is:
}

\textcolor{black}{
\begin{equation}
  E = \frac{P \cdot K / v_d}{K \cdot \alpha(K) + 1}
  \quad [\text{J/tok}],
\end{equation}
}

\textcolor{black}{
where the numerator is the drafting energy per round and the denominator is the expected number of accepted tokens.
}


The purpose of \namex is configuration-level exploration. \textcolor{black}{
For each configuration (draft model, device, and target model), \namex measures $v_d$, $\alpha(K)$, and $P$, and evaluates $G(K)$, $\eta_{\text{cost}}$, and $E$ using the analytical model above. Because $\eta_{\text{cost}}$ is independent of drafting speed and $E$ depends only on $v_d$ and $P$, configurations can be compared by profiling edge-side behavior and measuring draft–target alignment, while treating cloud verification latency as a parameter.
}

\section{Evaluation}
\label{section:evaluation}

We evaluate the effectiveness of distributed speculative decoding across heterogeneous edge platforms by profiling a comprehensive set of draft model configurations and speculative lengths.
Our analysis proceeds along three axes---\emph{goodput}, \emph{verification cost efficiency}, and \emph{energy efficiency}---and culminates in a unified selection framework that maps deployment constraints to recommended configurations.

\paragraph{Experimental Setup}
We deploy draft models on three representative edge platforms: Raspberry Pi~4B (RPi~4B, Cortex-A72, 8\,GB RAM), Raspberry Pi~5 (RPi~5, Cortex-A76, 8\,GB RAM), and NVIDIA Jetson AGX Orin (64\,GB unified memory, Ampere GPU).
All draft models are served via \texttt{llama.cpp} in GGUF format with quantization variants ranging from Q4\_K\_M to Q8\_0.
Two target models are considered: \textbf{Llama-3.1-70B} and \textbf{Qwen3-32B}, both hosted on a remote cloud verifier. We use the instruction prompts from the Databricks Dolly 15K dataset \cite{DatabricksBlog2023DollyV2} as input prompts. The dataset contains 15,011 instruction-following records created by Databricks contributors.

\subsection{Goodput Analysis}\label{subsec:goodput}

\begin{table}[t]
\centering
\footnotesize
\caption{Acceptance rate ($\alpha$) of draft models against each target model with speculative length of 5. Higher values indicate stronger alignment between draft and target distributions.}
\label{tab:acceptance_rate}
\begin{tabular}{lclc}
\toprule
\multicolumn{2}{c}{\textbf{Target: Llama-3.1-70B}} & \multicolumn{2}{c}{\textbf{Target: Qwen3-32B}} \\
\cmidrule(lr){1-2} \cmidrule(lr){3-4}
\textbf{Draft Model} & $\boldsymbol{\alpha}$ & \textbf{Draft Model} & $\boldsymbol{\alpha}$ \\
\midrule
Llama-3.2-1B          & 0.462 & Qwen3-0.6B & 0.378 \\
Llama-3.2-1B-Instruct & 0.546 & Qwen3-1.7B & 0.466 \\
Llama-3.2-3B-Instruct & 0.572 & Qwen3-4B   & 0.487 \\
Llama-3.1-8B          & 0.622 & Qwen3-8B   & 0.522 \\
\bottomrule
\end{tabular}
\end{table}

\begin{figure}[!t]
    \centering
    \includegraphics[width=\linewidth]{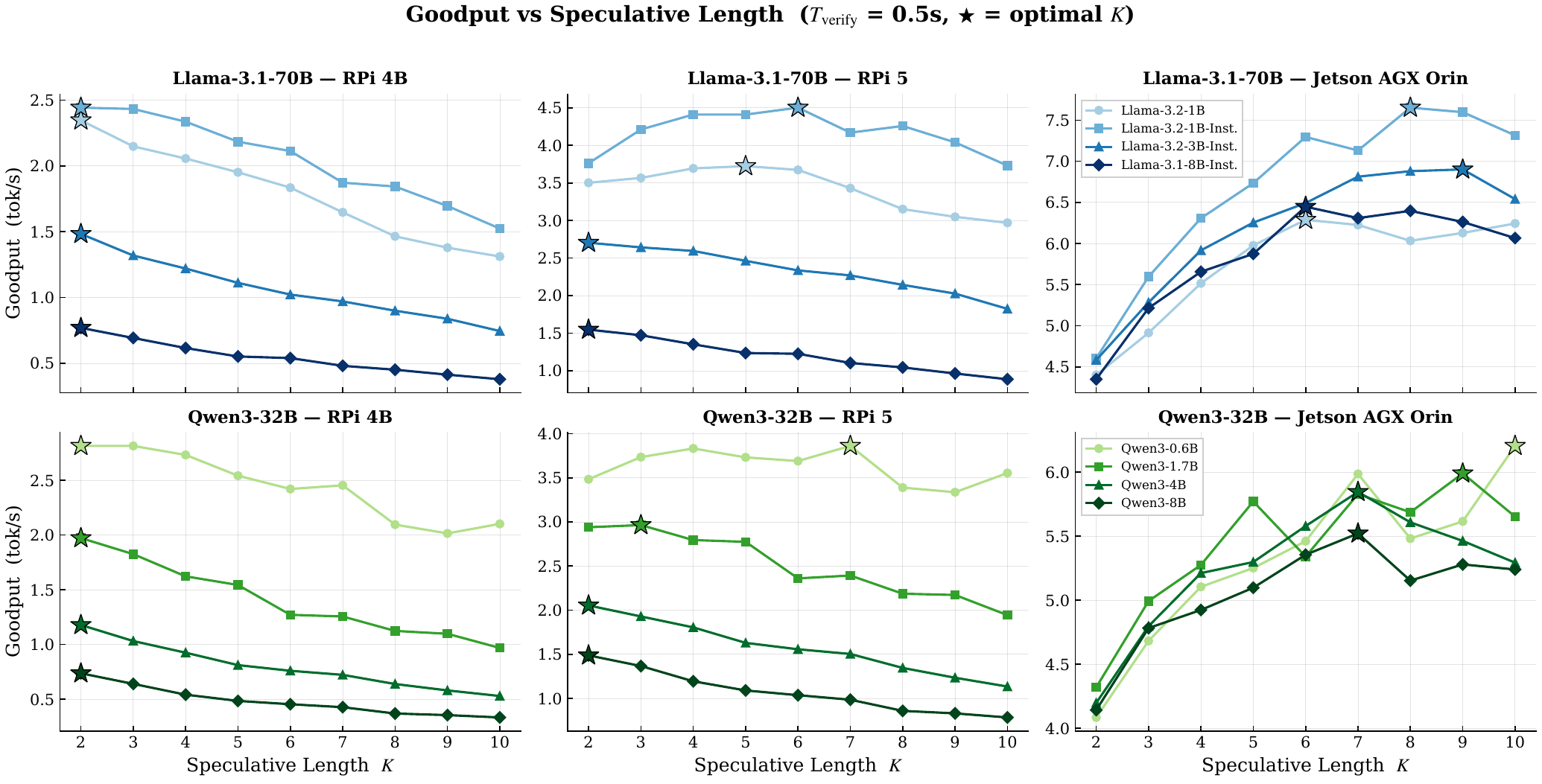}
     \vspace{-0.2in}
    \caption{Goodput vs. speculative length across devices and models.}
    \label{fig:goodput_spec_len}
\end{figure}

\subsubsection{Results}

Fig.~\ref{fig:goodput_spec_len} reveals that the optimal speculative length $K^*$ is
not a universal constant but varies significantly with both device speed and draft model size.
On the slowest device (RPi~4B), $K^*=2$ almost universally, as the high per-token drafting
cost ($1/v_d$) penalises additional speculation. In contrast, the Jetson AGX Orin
pushes $K^*$ to 5--8 for Llama and 4.5-7 for Qwen family, since its fast drafting makes the incremental cost of extra tokens negligible relative to the fixed verification latency $T_{\text{verify}}$, which is effectively amortised over more candidates. \textcolor{black}{$T_{\text{verify}}$ has been carefully selected based on historical experiments where the considered target models have been observed taking on average 0.5s to verify tokens, and it can vary for different target models and underlying hardware. Moreover, within each device, smaller draft models (e.g., Llama-3.2-1B, Qwen3-0.6B) consistently favour higher $K^*$ than their larger counterparts}, following the
same cost--benefit logic. On fast devices the goodput peaks are broad,
meaning the exact choice of $K$ is forgiving; on slow devices the curves are nearly flat,
making $K$ inconsequential. These observations suggest a practical guideline: set $K=2$ on
resource-constrained platforms and $K\geq5$ on GPU-class accelerators, with the precise value
tuned per draft--target pair via a lightweight sweep.

\begin{figure}[t]
    \centering
    \includegraphics[width=\linewidth]{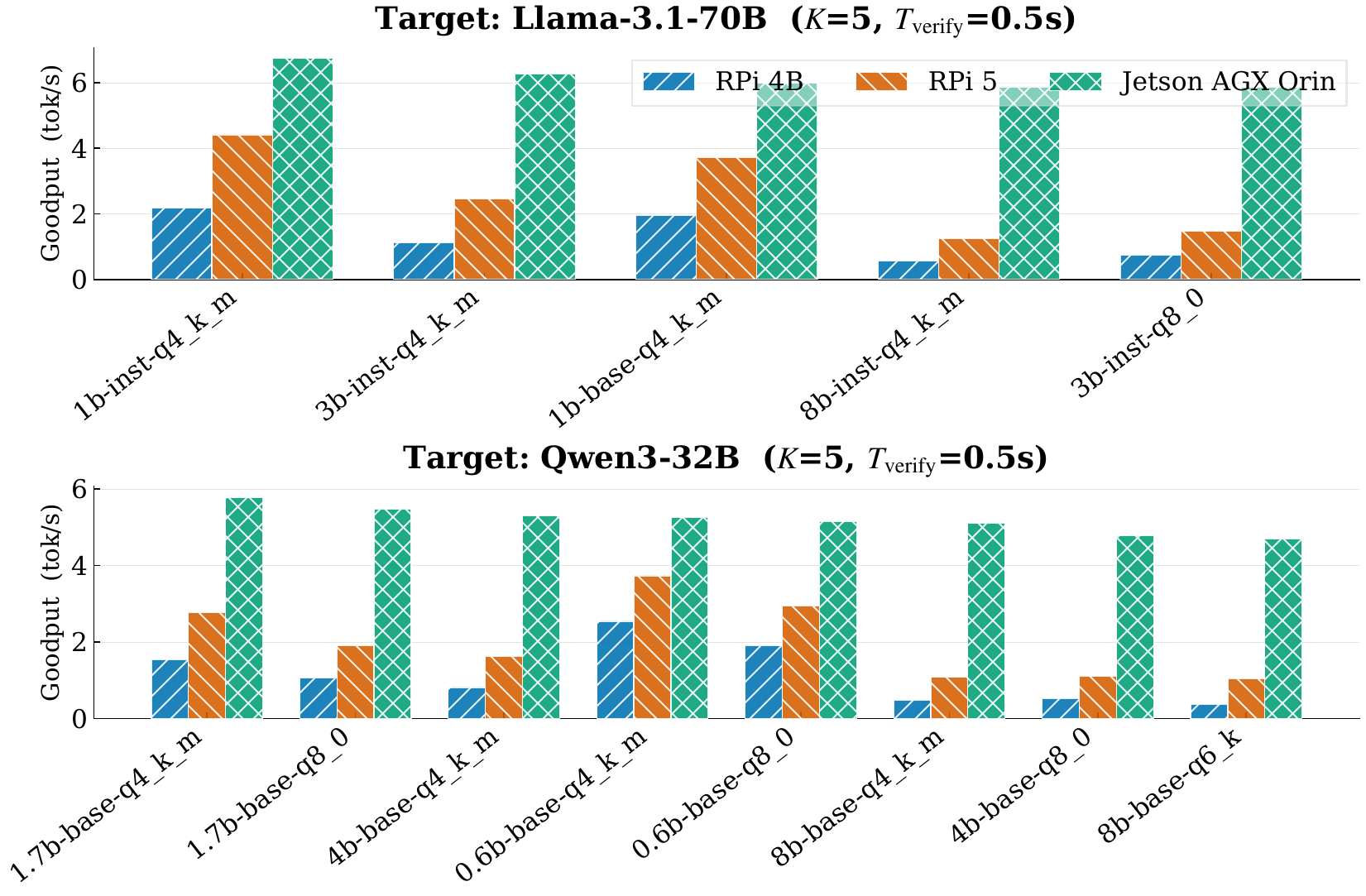}
    \vspace{-0.2in}
    \caption{Verified token generation speed.}
    \label{fig:goodput}
    \vspace{-0.15in}
\end{figure}

\paragraph{Goodput results.}
Fig.~\ref{fig:goodput} reports the goodput for Llama- and Qwen-family draft models across three devices.
In both families, the smallest draft models achieve the highest goodput:
Llama-3.2-1B (Q4\_K\_M) reaches approximately 6.5 tok/s on the Jetson AGX Orin, while
Qwen3-0.6B (Q4\_K\_M) reaches roughly 5.8 tok/s.
Scaling up the draft model consistently degrades goodput despite improving $\alpha$;
for instance, moving from Llama-3.2-1B to 3.1-8B raises $\alpha$ from 0.46 to 0.62
yet cuts goodput by more than $4\times$ on the Jetson.
A notable observation is that the Jetson AGX Orin does not outperform the Raspberry Pi
devices by as wide a margin as its raw drafting speed would suggest.
While the Jetson drafts tokens $6.5$--$16.2\times$ faster than the RPi~5, its goodput
advantage is only $1.5$--$2\times$.
This is because the verification latency $T_{\mathrm{verify}}$ is shared across all
devices---once the draft tokens are produced quickly, the system must still wait for the
remote server to verify them.
As $T_{\mathrm{verify}}$ dominates the round-trip time, additional drafting speed yields
diminishing returns on goodput.
On the RPi~4B, all models above 1B fall below 1 tok/s, rendering them impractical for
interactive use.
These results reveal a consistent structural trend across both model families:
\emph{on edge devices, goodput is governed by the interplay between drafting speed and
verification latency}.

\subsection{Verification Cost Efficiency}\label{subsec:cost}

\subsubsection{Pricing Data}
We adopt publicly listed inference pricing from major providers as representative verification costs.
For the \textbf{Llama-3.1-70B} target, we use the Fireworks AI serverless tier for models exceeding 16B parameters, priced at \$0.90 per 1M tokens~\cite{fireworks_pricing}.
For the \textbf{Qwen3-32B} target, we use Groq's on-demand pricing at \$0.59 per 1M tokens~\cite{groq_pricing}.

\subsubsection{Results}

\begin{figure}[t]
    \centering
    \includegraphics[width=\linewidth]{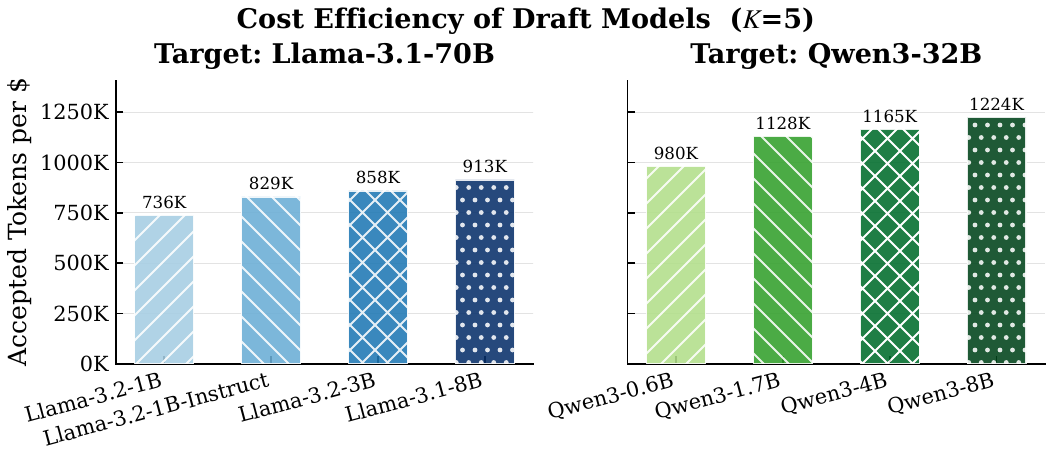}
     \vspace{-0.3in}
    \caption{Cost efficiency of draft models.}
    \label{fig:cost}
    \vspace{-0.15in}
\end{figure}

Fig.~\ref{fig:cost} shows that cost efficiency increases monotonically with draft model capacity in both families, consistent with Eq.~\eqref{eq:cost_efficiency}.
The largest drafts---Llama-3.1-8B (913K tokens/\$, $\alpha=0.62$) and Qwen3-8B (1224K tokens/\$, $\alpha=0.52$)---outperform their smallest counterparts by around 19.4\% and 24.9\%, respectively, driven entirely by higher acceptance rates.
The Qwen configurations achieve uniformly higher cost efficiency due to a lower verification price (\$0.59/M vs.\ \$0.90/M for Llama).
Combined with the goodput results in Section~\ref{subsec:goodput}, this reveals a fundamental Pareto tradeoff:
\emph{smaller draft models maximize goodput} by maximizing $v_d$, while \emph{larger draft models maximize cost efficiency} by maximizing $\alpha$.
The optimal choice therefore depends on whether the deployment prioritizes latency or monetary cost, a point we revisit in Section~\ref{subsec:selection}.

\subsection{Energy Efficiency}\label{subsec:energy}

\begin{figure}[t]
    \centering
    \includegraphics[width=\linewidth]{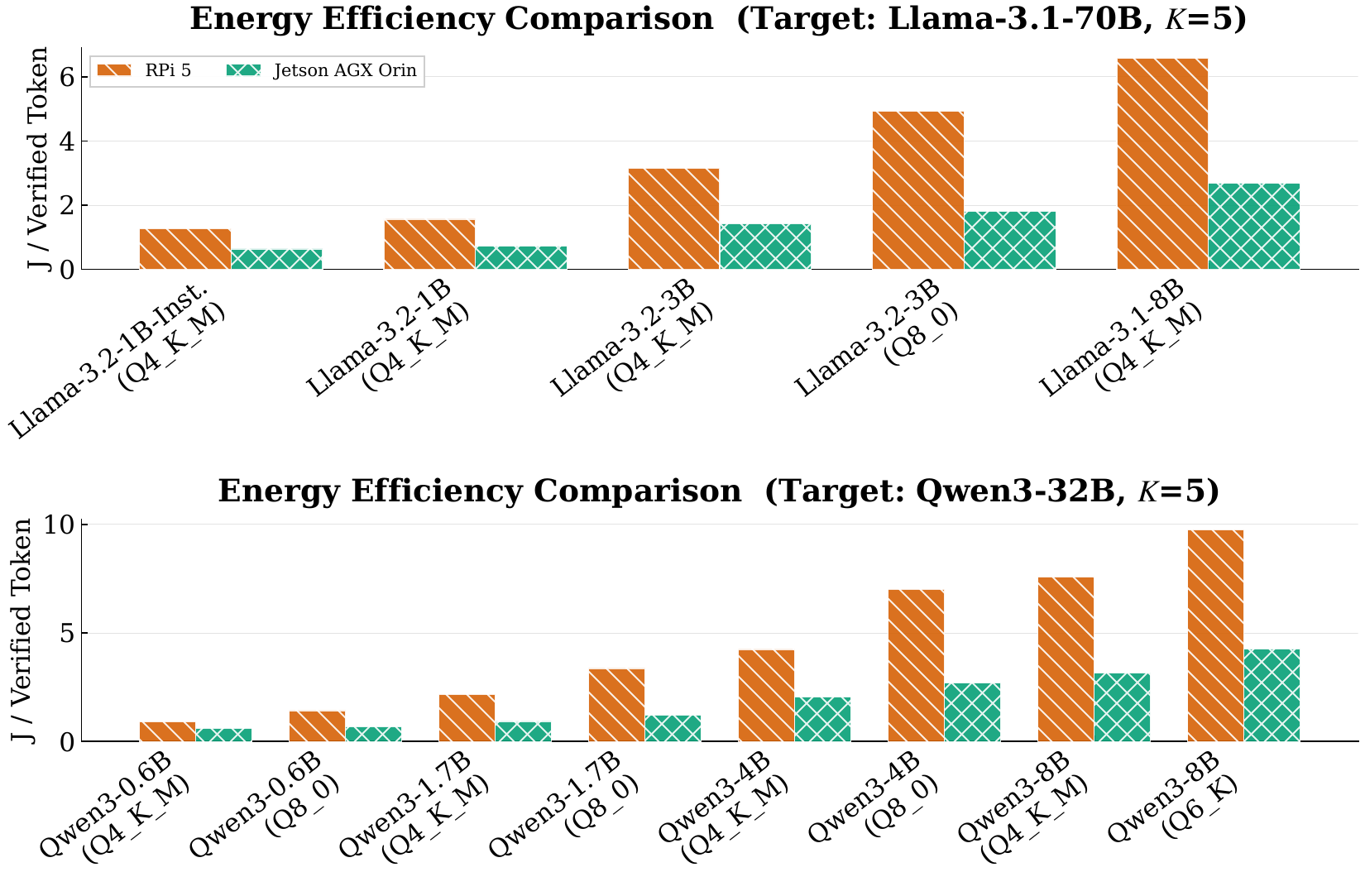}
     \vspace{-0.25in}
    \caption{Energy efficiency of draft models.}
    \label{fig:energy_bar}
    \vspace{-0.1in}
\end{figure}

\paragraph{Energy comparison results.}
Fig.~\ref{fig:energy_bar} shows that the edge device's energy per verified token escalates steeply with draft model size on both platforms \footnote{Real-time power monitoring is much less accessible on the Raspberry Pi 4B than on the Raspberry Pi 5, which provides more practical platform-level support for power measurement. Therefore we report power results only for the Raspberry Pi 5.}. The smallest drafts---Llama-3.2-1B-Instruct Q4\_K\_M (0.63\,J/tok on Jetson, 1.28\,J/tok on RPi~5) and Qwen3-0.6B Q4\_K\_M (0.60\,J/tok, 0.91\,J/tok)---achieve the best energy efficiency across all configurations.
Scaling to the largest drafts degrades efficiency by $4.17$--$5.14\times$ in the Llama family (Llama-3.1-8B: 2.67\,J/tok on Jetson, 6.58\,J/tok on RPi~5) and up to $10.75\times$ in the Qwen family (Qwen3-8B Q6\_K: 4.25\,J/tok on Jetson, 9.73\,J/tok on RPi~5).
The degradation is consistently less severe on the Jetson, whose higher compute throughput better amortizes the fixed power overhead.

\begin{figure}[t]
    \centering
    \includegraphics[width=\linewidth]{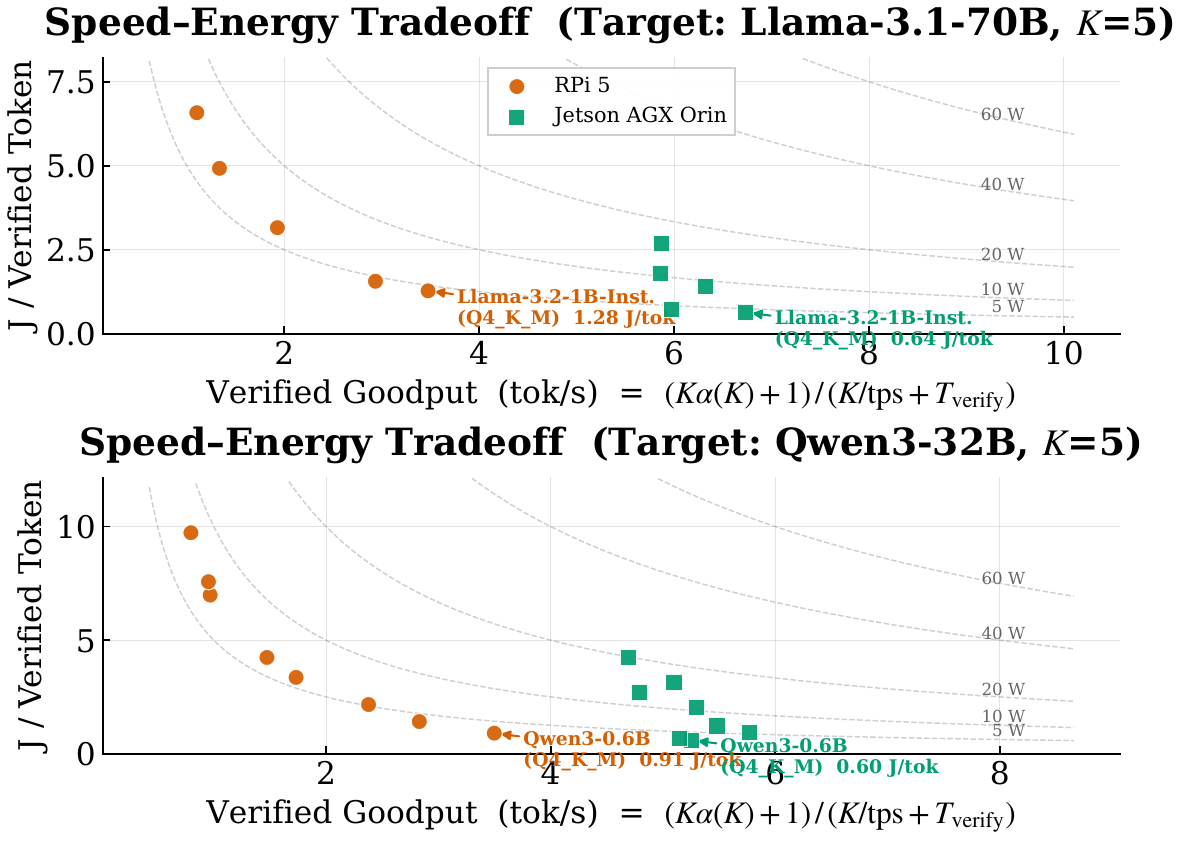}
     \vspace{-0.2in}
    \caption{Energy efficiency and \ speed (goodput) comparison.}
    \label{fig:energy_scatter}
     \vspace{-0.15in}
\end{figure}

\paragraph{Speed--energy Pareto front.}
Fig.~\ref{fig:energy_scatter} visualizes the speed--energy tradeoff with iso-power curves at 15\,W, 20\,W, 40\,W, and 60\,W.
In both model families, the smallest-draft model with lowest quantization bit-width on Jetson configurations occupy the Pareto-optimal corner (high goodput, low energy), while all RPi~5 configurations are Pareto-dominated by their Jetson counterparts.
This architectural advantage stems from the Jetson's GPU-accelerated inference, which exploits massive parallelism in matrix operations to deliver substantially higher throughput per watt, whereas the RPi~5's CPU-only execution serializes these operations across a small number of cores, resulting in longer active inference time and proportionally greater energy expenditure per token.

\subsection{Optimal Configuration Selection across the Three-Dimensional Space}
\label{subsec:selection}

Sections~4.1--4.3 examined goodput, cost, and energy in isolation while fixing the speculative length at $K=5$.
We now treat \emph{all three} configuration knobs---\textbf{draft-model variant} $M$, \textbf{quantisation level} $Q$, and \textbf{speculative length $K$}---as a joint search space and ask:
for a given LLM family and edge device, which triple $(M, Q, K)$ optimises each objective. \footnote{Considering the power supply on the server is continuous, we only consider the power consumption and optimization on edge device side}

\paragraph{Methodology.}
For every (target, device) pair, we enumerate all feasible $(M, Q, K)$ triples with $K \in \{2,\dots,10\}$ and evaluate three metrics:
(i)~goodput $G$ (Eq.~1),
(ii)~cost efficiency $\eta_{\mathrm{cost}}$ (Eq.~2), and
(iii)~energy per accepted token $E$ (Eq.~3). 
Table~\ref{tab:optimal-config} reports the configuration that optimises each metric, together with \emph{all three} metric values so that the trade-offs are directly visible. \textcolor{black}{As the $K$ and acceptance rate ($\alpha$) have an intricate relation, for each $M, Q, K$, edge device, and target model combination, we computed tailored $\alpha(K)$}.

\begin{table}[t]
\centering
\footnotesize
\setlength{\tabcolsep}{4pt}
\renewcommand{\arraystretch}{1.05}
\caption{Recommended configurations under different optimisation objectives.
$G$: goodput (tok/s), $\eta_{\mathrm{cost}}$: cost efficiency (K\,tokens/\$), $E$: energy per accepted token (J/tok).
Best values per (target, device) group are \textbf{bolded}.}
\label{tab:optimal-config}
\begin{tabular}{lll l c r r r}
\toprule
\textbf{Target} & \textbf{Device} & \textbf{Objective} & \textbf{Configuration} & $K$ & $G$ & $\eta_{\mathrm{cost}}$ & $E$ \\
\midrule
\multirow{9}{*}{\rotatebox{90}{Llama-3.1-70B}}
 & \multirow{3}{*}{RPi 4B}
   & Max Goodput  & Llama-3.2-1B-Inst Q4   & 2  & \textbf{2.44} & 1334K          & --- \\
 & & Min Cost/tok & Llama-3.1-8B-Inst Q4   & 2  & 0.77          & \textbf{1401K} & --- \\
 & & Min Energy   & \multicolumn{5}{c}{\emph{no power data}} \\
\cmidrule(lr){2-8}
 & \multirow{3}{*}{RPi 5}
   & Max Goodput  & Llama-3.2-1B-Inst Q4   & 6  & \textbf{4.50} & 763K           & 0.84 \\
 & & Min Cost/tok & Llama-3.1-8B-Inst Q4   & 2  & 1.55          & \textbf{1401K} & 3.75 \\
 & & Min Energy   & Llama-3.2-1B-Inst Q4   & 2  & 3.76          & 1334K          & \textbf{0.48} \\
\cmidrule(lr){2-8}
 & \multirow{3}{*}{Jetson}
   & Max Goodput  & Llama-3.2-1B-Inst Q4   & 8  & \textbf{7.65} & 623K           & 0.85 \\
 & & Min Cost/tok & Llama-3.1-8B-Inst Q4   & 2  & 4.35          & \textbf{1401K} & 1.74 \\
 & & Min Energy   & Llama-3.2-1B-Inst Q4   & 2  & 4.60          & 1334K          & \textbf{0.39} \\
\midrule
\multirow{9}{*}{\rotatebox{90}{Qwen3-32B}}
 & \multirow{3}{*}{RPi 4B}
   & Max Goodput  & Qwen3-0.6B Q4          & 2  & \textbf{2.81} & 1801K          & --- \\
 & & Min Cost/tok & Qwen3-8B Q4            & 2  & 0.74          & \textbf{2048K} & --- \\
 & & Min Energy   & \multicolumn{5}{c}{\emph{no power data}} \\
\cmidrule(lr){2-8}
 & \multirow{3}{*}{RPi 5}
   & Max Goodput  & Qwen3-0.6B Q4          & 7  & \textbf{3.86} & 828K           & 0.90 \\
 & & Min Cost/tok & Qwen3-8B Q4            & 2  & 1.49          & \textbf{2048K} & 3.86 \\
 & & Min Energy   & Qwen3-0.6B Q4          & 2  & 3.48          & 1801K          & \textbf{0.41} \\
\cmidrule(lr){2-8}
 & \multirow{3}{*}{Jetson}
   & Max Goodput  & Qwen3-0.6B Q4          & 10 & \textbf{6.21} & 633K           & 0.93 \\
 & & Min Cost/tok & Qwen3-8B Q4            & 2  & 4.14          & \textbf{2048K} & 1.88 \\
 & & Min Energy   & Qwen3-0.6B Q4          & 2  & 4.08          & 1801K          & \textbf{0.33} \\
\bottomrule
\end{tabular}
\end{table}

\paragraph{Observation 1: Goodput favours small, fast drafters with device-dependent $K^*$.}
Across both families, the smallest Q4 quantised draft model---Llama-3.2-1B-Instruct for Llama-70B and Qwen3-0.6B for Qwen3-32B---delivers the highest goodput on every device.
The optimal speculative length $K^*$, however, scales with device speed:
on the RPi~4B the fixed verification overhead $T_{\mathrm{verify}}$ already dominates at $K=2$;
on the RPi~5 the sweet spot rises to $K^*=6$--7;
and on the Jetson it climbs to $K^*=8$--10.
Comparing the Jetson's goodput-optimal $G=7.65$\,tok/s (Llama) with the RPi~4B's $G=2.44$\,tok/s reveals only a $3.1\times$ advantage despite a ${\sim}22\times$ raw speed gap, confirming that $T_{\mathrm{verify}}$ compresses the goodput range.
The cost of this goodput-maximising strategy is visible in the $\eta_{\mathrm{cost}}$ column: on the Jetson it drops to 623K\,tok/\$---less than half the cost-optimal 1401K\,tok/\$.

\paragraph{Observation 2: Cost efficiency is device-independent and always peaks at $K=2$ with the largest drafter.}
Since $\eta_{\mathrm{cost}} = (\alpha(K)+1/K)/p$ depends only on the acceptance rate and verifier price, the cost-optimal configuration is identical across all three devices:
8B-Instruct Q4 at $K=2$ for Llama-70B and Qwen3-8B Q4 at $K=2$ for Qwen3-32B.
The key is the \emph{bonus-token effect}: every verification round produces one auto-regressive output token regardless of how many drafted tokens are rejected.
At $K=2$ this free token contributes $1/K=0.5$ to the per-verified-token yield, whereas at $K=10$ it contributes only $0.1$.
Combined with the high acceptance rate of 8B drafters at short sequences ($\alpha(2)\approx 0.76$), $K=2$ yields $\alpha(2)+1/2=1.26$ accepted tokens per verified token---the maximum across the entire search space.
The trade-off is stark: on the RPi~5 the cost-optimal configuration delivers only $G=1.55$\,tok/s ($2.9\times$ slower than the goodput-optimal 4.50\,tok/s) and $E=3.75$\,J/tok ($7.8\times$ worse energy than the energy-optimal 0.48\,J/tok).

\paragraph{Observation 3: Energy efficiency mirrors goodput's model choice but universally locks to $K=2$.}
The energy-optimal draft model is the same small, fast model that maximises goodput---1B-Instruct Q4 for Llama, 0.6B Q4 for Qwen---because these draw the least power and minimise drafting time.
Yet the optimal speculative length is universally $K=2$, diverging from goodput's device-dependent $K^*$.
Since $E$ only counts local drafting energy, setting $K=2$ minimises the joules spent per round while the bonus-token effect keeps yield high.
On the Jetson the energy-optimal $E=0.39$\,J/tok (Llama) is $17\%$ lower than the RPi~5's 0.48\,J/tok despite higher idle power, because the Jetson completes each 2-token draft in a fraction of the time.
Notably, the energy-optimal configuration still achieves reasonable goodput---$G=4.60$\,tok/s on the Jetson (60\% of the goodput-optimal 7.65)---making it an attractive default for battery-constrained deployments where moderate latency is acceptable.

\paragraph{Key trade-offs.}
\begin{itemize}[nosep]
  \item \textbf{Model size}: Goodput and energy prefer the \emph{smallest} drafter (maximum $v_d$, minimum $P$), while cost prefers the \emph{largest} (maximum $\alpha$). On the RPi~5, switching from the cost-optimal 8B to the goodput-optimal 1B improves throughput by $2.9\times$ and energy by $7.8\times$, but sacrifices 46\% cost efficiency.
  \item \textbf{Speculative length $K$}: Goodput benefits from longer speculation on fast devices (more tokens amortise $T_{\mathrm{verify}}$), while both cost and energy favour $K=2$ (the bonus-token effect dominates). Reducing $K$ from the goodput-optimal $K^*$ to~2 costs up to 40\% goodput but can halve energy per token.
\end{itemize}
These results demonstrate that profiling-based configuration selection across the joint $(M, Q, K)$ space is essential: no single fixed setting can simultaneously optimise throughput, cost, and energy.

\section{Conclusion}

\namex is a framework for configuration selection in distributed speculative LLM serving across cloud and edge. By modeling heterogeneous edge devices and speculative decoding dynamics, \namex enables rapid exploration of trade-offs among edge model variants, quantisation levels, and speculative lengths without repeated system deployment. Our analysis of the joint $(M,Q,K)$ configuration space reveals that the three optimisation objectives impose fundamentally conflicting pressures: goodput favours small, fast drafters at device-dependent $K^*$ that amortises verification latency, while both cost and energy efficiency converge to $K{=}2$ due to the dominant bonus-token effect---though cost selects the largest drafter to maximise acceptance rate whereas energy selects the smallest to minimise power draw. These structural conflicts confirm that no single fixed configuration can simultaneously optimise throughput, cost, and energy, underscoring the necessity of profiling-based selection. The framework complements deployment-focused speculative decoding systems by supporting principled design-space exploration in disaggregated AI inference infrastructures.

\section{Acknowledgements}

\par This work was supported by a research grant from the Department for the Economy, Northern Ireland (grant agreement USI-226), Virginia Tech College of Engineering (grant Major Grants Initiative Program), and by the National Science Foundation (grant No. 2315851).

\bibliographystyle{ACM-Reference-Format}
\bibliography{references}

@inproceedings{tianclone,
  author = {Tian, Chunlin and Qin, Xinpeng and Tam, Kahou and Li, Li and Wang, Zijian and Zhao, Yuanzhe and Zhang, Minglei and Xu, Chengzhong},
  title = {CLONE: customizing LLMs for efficient latency-aware inference at the edge},
  year = {2025},
  isbn = {978-1-939133-48-9},
  publisher = {USENIX Association},
  address = {USA},
  booktitle = {Proceedings of the 2025 USENIX Conference on Usenix Annual Technical Conference},
  articleno = {34},
  numpages = {23},
  location = {Boston, MA, USA},
  series = {USENIX ATC '25}
}

@article{han2025privacy,
  title={A privacy-preserving and trustworthy inference framework for LLM-IoT integration via hierarchical federated collaborative computing},
  author={Han, Chengzhuo and Yang, Tingting and Cui, Zhengqi and Sun, Xin},
  journal={IEEE Internet of Things Journal},
  year={2025},
  publisher={IEEE}
}

@inproceedings{yu2024edge,
  title={Edge-llm: Enabling efficient large language model adaptation on edge devices via unified compression and adaptive layer voting},
  author={Yu, Zhongzhi and Wang, Zheng and Li, Yuhan and Gao, Ruijie and Zhou, Xiaoya and Bommu, Sreenidhi Reddy and Zhao, Yang and Lin, Yingyan},
  booktitle={Proceedings of the 61st ACM/IEEE Design Automation Conference},
  pages={1--6},
  year={2024}
}

@inproceedings{xiao2023smoothquant,
  title={Smoothquant: Accurate and efficient post-training quantization for large language models},
  author={Xiao, Guangxuan and Lin, Ji and Seznec, Mickael and Wu, Hao and Demouth, Julien and Han, Song},
  booktitle={International Conference on Machine Learning},
  pages={38087--38099},
  year={2023},
  organization={PMLR}
}

@inproceedings{frantar2023sparsegpt,
  title={Sparsegpt: Massive language models can be accurately pruned in one-shot},
  author={Frantar, Elias and Alistarh, Dan},
  booktitle={International Conference on Machine Learning},
  pages={10323--10337},
  year={2023},
  organization={PMLR}
}

@inproceedings{chen2025efficientqat,
  title={Efficientqat: Efficient quantization-aware training for large language models},
  author={Chen, Mengzhao and Shao, Wenqi and Xu, Peng and Wang, Jiahao and Gao, Peng and Zhang, Kaipeng and Luo, Ping},
  booktitle={Proceedings of the 63rd Annual Meeting of the Association for Computational Linguistics (Volume 1: Long Papers)},
  pages={10081--10100},
  year={2025}
}

@article{li2025adaptive,
  title={Adaptive model partitioning and pruning for collaborative DNN inference in mobile edge-cloud computing networks},
  author={Li, Hui and Li, Xiuhua and Fan, Qilin and He, Qiang and Wang, Xiaofei and Leung, Victor CM},
  journal={IEEE Transactions on Mobile Computing},
  year={2025},
  publisher={IEEE}
}

@inproceedings{leviathan2023fast,
  title={Fast inference from transformers via speculative decoding},
  author={Leviathan, Yaniv and Kalman, Matan and Matias, Yossi},
  booktitle={International Conference on Machine Learning},
  pages={19274--19286},
  year={2023},
  publisher={PMLR},
  address={Honolulu, Hawaii, USA}
}

@article{miao2023specinfer,
  title={Specinfer: Accelerating generative llm serving with speculative inference and token tree verification},
  author={Miao, Xupeng and Oliaro, Gabriele and Zhang, Zhihao and Cheng, Xinhao and Wang, Zeyu and Wong, Rae Ying Yee and Chen, Zhuoming and Arfeen, Daiyaan and Abhyankar, Reyna and Jia, Zhihao},
  journal={arXiv preprint arXiv:2305.09781},
  volume={1},
  number={2},
  pages={4},
  year={2023}
}

@article{chen2023accelerating,
  title={Accelerating large language model decoding with speculative sampling},
  author={Chen, Charlie and Borgeaud, Sebastian and Irving, Geoffrey and Lespiau, Jean-Baptiste and Sifre, Laurent and Jumper, John},
  journal={arXiv preprint arXiv:2302.01318},
  year={2023}
}

@inproceedings{li2025sled,
  title={Sled: A speculative llm decoding framework for efficient edge serving},
  author={Li, Xiangchen and Spatharakis, Dimitrios and Ghafouri, Saeid and Fan, Jiakun and Vandierendonck, Hans and John, Deepu and Ji, Bo and Nikolopoulos, Dimitrios S},
  booktitle={Proceedings of the Tenth ACM/IEEE Symposium on Edge Computing},
  pages={1--8},
  year={2025}
}

@inproceedings{sadhukhanmagicdec,
  title={MagicDec: Breaking the Latency-Throughput Tradeoff for Long Context Generation with Speculative Decoding},
  author={Sadhukhan, Ranajoy and Chen, Jian and Chen, Zhuoming and Tiwari, Vashisth and Lai, Ruihang and Shi, Jinyuan and Yen, Ian En-Hsu and May, Avner and Chen, Tianqi and Chen, Beidi},
  booktitle={International Conference on Learning Representations},
  year={2024}
}

@article{xu2024edgellm,
  title={Edgellm: Fast on-device llm inference with speculative decoding},
  author={Xu, Daliang and Yin, Wangsong and Zhang, Hao and Jin, Xin and Zhang, Ying and Wei, Shiyun and Xu, Mengwei and Liu, Xuanzhe},
  journal={IEEE Transactions on Mobile Computing},
  volume={24},
  number={4},
  pages={3256--3273},
  year={2024},
  publisher={IEEE}
}

@article{zhang2024edgeshard,
  title={EdgeShard: Efficient LLM inference via collaborative edge computing},
  author={Zhang, Mingjin and Shen, Xiaoming and Cao, Jiannong and Cui, Zeyang and Jiang, Shan},
  journal={IEEE Internet of Things Journal},
  year={2024},
  publisher={IEEE}
}

@misc{fireworks_pricing,
  author       = {{Fireworks AI}},
  title        = {Pricing -- {Fireworks AI}},
  year         = {2025},
  howpublished = {\url{https://fireworks.ai/pricing}},
  note         = {Serverless tier: \$0.90 / 1M tokens for models {>}16B parameters. Accessed: 2025-07-15}
}

@misc{groq_pricing,
  author       = {{Groq}},
  title        = {{GroqCloud} On-Demand Pricing},
  year         = {2025},
  howpublished = {\url{https://groq.com/pricing}},
  note         = {Qwen3-32B: \$0.29 / 1M input tokens, \$0.59 / 1M output tokens. Accessed: 2025-07-15}
}

@article{li2026wisp,
  title={WISP: Waste-and Interference-Suppressed Distributed Speculative LLM Serving at the Edge via Dynamic Drafting and SLO-Aware Batching},
  author={Li, Xiangchen and Fan, Jiakun and Wang, Qingyuan and Spatharakis, Dimitrios and Ghafouri, Saeid and Vandierendonck, Hans and John, Deepu and Ji, Bo and Butt, Ali R and Nikolopoulos, Dimitrios S},
  journal={arXiv preprint arXiv:2601.11652},
  year={2026}
}

@online{DatabricksBlog2023DollyV2,
    author    = {Mike Conover and Matt Hayes and Ankit Mathur and Jianwei Xie and Jun Wan and Sam Shah and Ali Ghodsi and Patrick Wendell and Matei Zaharia and Reynold Xin},
    title     = {Free Dolly: Introducing the World's First Truly Open Instruction-Tuned LLM},
    year      = {2023},
    url       = {https://www.databricks.com/blog/2023/04/12/dolly-first-open-commercially-viable-instruction-tuned-llm},
    urldate   = {2023-06-30}
}

\end{document}